# HIGH PRESSURE TRANSPORT STUDY OF NON-FERMI LIQUID BEHAVIOUR IN $U_2Pt_2In$ AND $U_3Ni_3Sn_4$

ANNE DE VISSER AND PEDRO ESTRELA

*Van der Waals-Zeeman Institute, University of Amsterdam,*
*Valckenierstraat 65, 1018 XE Amsterdam, The Netherlands*
*E-mail: devisser@science.uva.nl*

TAKASHI NAKA

*National Research Institute for Metals, 1-2-1 Sengen, Ibaraki 305-0047, Japan*
*E-mail: naka@nrim.go.jp*

The strongly correlated metals $U_2Pt_2In$ and $U_3Ni_3Sn_4$ show pronounced non-Fermi liquid (NFL) phenomena at ambient pressure. Here we review single-crystal electrical resistivity measurements under pressure ($p \leq 1.8$ GPa) conducted to investigate the stability of the NFL phase. For tetragonal $U_2Pt_2In$ ($I \parallel a$) we observe a rapid recovery of the Fermi-liquid $T^2$-term with pressure. The Fermi-liquid temperature varies as $T_{FL} \sim p-p_c$, where $p_c = 0$ is a critical pressure. The analysis within the magnetotransport theory of Rosch provides evidence for the location of $U_2Pt_2In$ at a $p = 0$ antiferromagnetic quantum critical point (QCP). In the case of cubic $U_3Ni_3Sn_4$ we find $T_{FL} \sim (p-p_c)^{1/2}$. The analysis provides evidence for an antiferromagnetic QCP in $U_3Ni_3Sn_4$ at a negative pressure $p_c = -0.04 \pm 0.04$ GPa.

## 1 Introduction

Materials with low-temperature properties that show strong departures from the standard Fermi-liquid behaviour nowadays attract considerable attention. Exemplary compounds to study this so-termed *non-Fermi liquid* (NFL) behaviour are found amongst dense Kondo *f*-electron systems. A mechanism of much general interest that may lead to NFL behavior is the proximity to a magnetic quantum critical point (QCP) [1]. Here a $T = 0$ K electronic instability leads to a phase transition between a strongly renormalised Fermi liquid and a magnetically ordered phase. The *quantum phase transition* can occur spontaneously or may be induced by tuning the magnetic ordering temperature to 0 K by an external parameter, such as hydrostatic or chemical pressure. Magnetic quantum phase transitions in correlated metals are often discussed in terms of the "weak-coupling" Hertz-Millis model, in which the QCP is treated as a magnetic instability of the Fermi surface [1]. The Kondo-screened heavy quasiparticles undergo an antiferromagnetic (AF) spin-density wave transition. NFL behaviour is due to Bragg diffraction of the electrons off a critical spin-density wave. Recently, Coleman and coworkers [2] proposed an alternative "strong-coupling" local-moment model, which starts from the magnetic side and treats the metal as a local-moment antiferromagnet, that becomes disordered when the effective Kondo temperature, $T_K^*$, is large enough to form the dense Kondo state. At the QCP the Kondo-bound states decompose and $T_K^*$ vanishes.

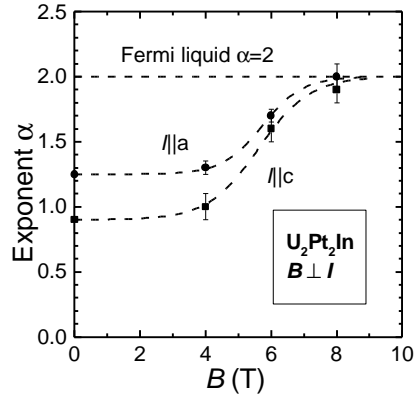

**Figure 1**. Field dependence of the resistivity exponent *a* of $U_2Pt_2In$. Dashed lines are to guide the eye.

Magnetotransport under pressure is an excellent experimental tool to explore the NFL phase. By comparing the experimental data to the predictions of a new magnetotransport theory by Rosch [3] for metals close to an AF QCP, a stringent test of the Hertz-Millis scenario can be made. One of the major assets of the Rosch theory is that it takes into account the effect of disorder. Considering that spin fluctuations are destroyed at the temperature scale $G$, where $G$ is typically of the order of $T_K$ or the coherence temperature $T_{coh}$, the following scaling form of the resistivity is derived [3]: $r = r_0 + \Delta r = r_0 + T^{3/2} f(T/r_0, (d-d_c)/r_0, B/r_0^{3/2})$. Here $r_0$ is the residual resistivity, $B$ the magnetic field and $d-d_c$ the distance to the QCP in the non-ordered (paramagnetic) side of the phase diagram, with $d_c$ the critical control parameter. The scaling form of the resistivity allows one to delineate the NFL ($\Delta r \sim T$ and $\Delta r \sim T^{3/2}$) and Fermi-liquid ($\Delta r \sim T^2$) regimes as a function of the distance to the QCP (here measured by the pressure $p$) and the amount of disorder in the material, $x \sim r_0/r_{RT}$. For instance, $T_{FL}$, i.e. the temperature below which $\Delta r \sim T^2$, varies initially as $T_{FL} = a_1 (p-p_c)$, with a cross-over to $T_{FL} = a_2 (p-p_c)^{1/2}$ at higher distances, where $p_c$ is the pressure at the QCP. Predictions for the NFL resistivity exponents in the local-moment model [2] are not available yet. However, the model predicts a discontinuity in the Hall constant at the QCP when the heavy-fermions decompose.

In this paper we review electrical resistivity measurements under pressure ($p \leq 1.8$ GPa), conducted to investigate the stability of the NFL phase in the heavy-electron compounds $U_2Pt_2In$ and $U_3Ni_3Sn_4$. $U_2Pt_2In$ belongs to the group of $U_2T_2X$ compounds, where T is a transition metal and X is In or Sn [4]. It does not show magnetic order and is the heaviest compound in this series ($c/T = 0.41$ J/(mol-U)K$^2$ at $T = 1$ K). $U_3Ni_3Sn_4$ belongs to the family of isostructural cubic stannides $U_3T_3Sn_4$, where T = Ni, Cu, Pt or Au. It is a moderate heavy-electron compound ($c/T = 0.09$ J/(mol-U)K$^2$) [5].

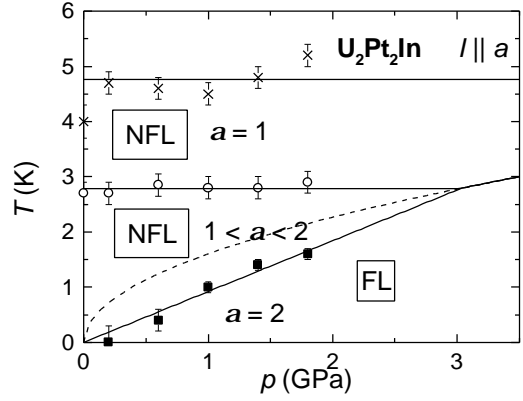

**Figure 2**. Pressure dependence of $T_{FL}$ (closed squares) and of the temperature range in which $r \sim T$ (between (o) and (x)) for $U_2Pt_2In$ ($I \parallel a$). The solid lines delineate NFL and FL regimes with exponents $\alpha$. The dashed line shows the $T_{FL} = a_2 (p-p_c)^{1/2}$ dependence. After Ref.8.

## 2    Resistivity of $U_2Pt_2In$ under pressure

$U_2Pt_2In$ shows robust NFL behaviour [6,7]. The NFL properties are summarised by: (i) the specific heat varies as $c(T) \sim -T\ln(T/T_0)$ over almost two decades of temperature ($T = 0.1$-6 K), (ii) the magnetic susceptibility shows a weak maximum at $T_m = 8$ K for $B \parallel c$ (tetragonal structure), while it increases as $T^{0.7}$ when $T \rightarrow 0$ for $B \parallel a$, and (iii) the electrical resistivity obeys a power law $T^\alpha$ with **a** = 1.25±0.05 ($T < 1$ K) and 0.9±0.1 ($T \rightarrow 0$), for the current along the $a$ and $c$ axis, respectively. Moderate magnetic fields strongly influence the NFL state. The Fermi-liquid value **a** = 2 is recovered in an applied magnetic field of ~ 8 T as shown in Fig. 1 [7]. Muon spin relaxation experiments exclude the absence of (weak) static magnetic order down to 0.05 K [7]. We have carried out high-pressure transport measurements on single-crystalline $U_2Pt_2In$ [8]. For $I \parallel c$ $r(T)$ increases with pressure and develops a relative minimum at low temperatures ($T_{min}$~ 4.8 K at 1.8 GPa). In the following, we concentrate on the low-temperature data for $I \parallel a$.

A detailed analysis has been presented in Ref.8. In Fig. 2 we show how the results compare to the transport theory of Rosch [3] for materials with considerable disorder (for our $U_2Pt_2In$ crystal $x \sim 0.6$). The different NFL and FL regimes are deduced by fitting the resistivity under pressure to a $T^2$ term at the lowest temperatures and a term linear in $T$ at higher temperatures. Fig. 2 shows that the data are consistent with $T_{FL}$ being a linear function of pressure with $p_c = 0$. The cross-over to a $T_{FL} = a_2 (p-p_c)^{1/2}$ dependence is expected near 3.0 GPa. The $\Delta r \sim T$ region is predicted to occur in the reduced temperature range $x < T/\boldsymbol{G} < x^{1/2}$ ($x < 1$). From Fig. 2 we extract that the $\Delta r \sim T$ region is found in the temperature range 2.8-4.7 K, from which it follows $x = 0.34$ and $\boldsymbol{G} = 8.1$ K. The agreement between the calculated value $x = 0.34$ and the experimental value $x \sim 0.6$ is, given the rather simple data treatment, satisfactory. The temperature-pressure diagram presented in Fig. 2 is consistent with the scaling diagram for the resistivity presented by Rosch. Hence, we conclude that the compound $U_2Pt_2In$ has an AF QCP at a critical pressure $p_c = 0$.

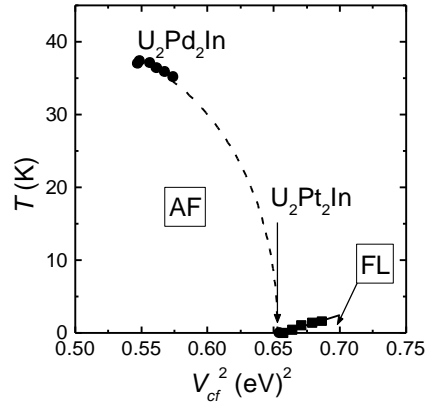

**Figure 3**. Tentative Doniach-type diagram for $U_2Pt_2In$ (squares: $T_{FL}(p)$) and $U_2Pd_2In$ (circles: $T_N(p)$). AF= antiferromagnetic order, FL = Fermi-liquid regime. The lines serve to guide the eye.

It is of interest to compare the pressure effects on $U_2Pt_2In$ and the isoelectronic compound $U_2Pd_2In$. $U_2Pd_2In$ is an antiferromagnet with $T_N = 37.5 \pm 0.5$ K, which decreases under pressure to $T_N = 35.2 \pm 0.5$ K at $p = 1.8$ GPa [7]. In Fig. 3 we show a tentative Doniach-like diagram for the compounds $U_2Pt_2In$ and $U_2Pd_2In$ under pressure. Here the relative increase of $V_{cf}$, i.e. the hybridization matrix element for the total conduction electron hybridization at the $f$ atom, was calculated using a simple model (see Ref.7). With a compressibility value $\boldsymbol{k} = 6.82 \times 10^{-3}$ GPa$^{-1}$, we calculate for $U_2Pt_2In$ that $V_{cf}$ increases by 2.3% in the pressure range 0-1.8 GPa. Thus by applying a moderate pressure of 1.8 GPa, $U_2Pt_2In$ is shifted considerably into the non-magnetic region of the Doniach-like diagram. A similar calculation for $U_2Pd_2In$ results in an increase of $V_{cf}$ of 2.4% at 1.8 GPa. According to Fig. 3, a very rough estimate of the pressure needed to tune $U_2Pd_2In$ to a QCP is ~ 7 GPa.

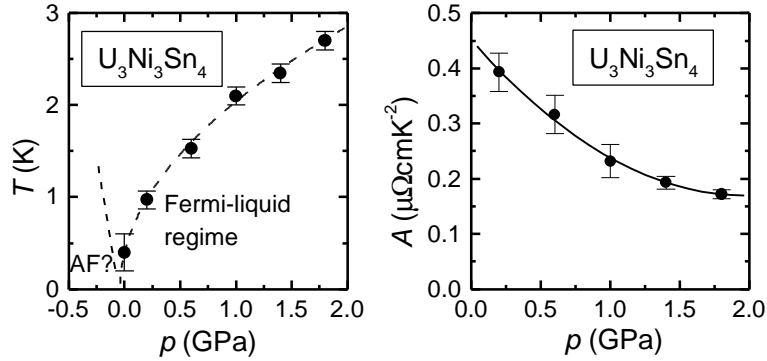

**Figure 4**. $T_{FL}$ and the coefficient A as function of pressure for $U_3Ni_3Sn_4$ (After Ref.10).

## 3 Resistivity of $U_3Ni_3Sn_4$ under pressure

For $U_3Ni_3Sn_4$, evidence for the proximity to an AF instability is predominantly provided by specific-heat experiments [9]. Data taken on a single-crystalline sample in the temperature range $T = 0.3$-$5$ K, revealed the presence of a NFL electronic term of the form $c_{el}/T \sim g_0 - aT^{1/2}$. Such a $aT^{1/2}$ correction to the standard Fermi-liquid coefficient has been predicted for a 3D AF QCP by Millis. Subsequent measurements, showed that the electronic specific heat below $T = 0.4$ K is best described by the modified Fermi-liquid expression $c_{el}/T \sim g_0 + dT^3 \ln(T/T^*)$, with $g_0 = 0.130$ J/(mol-U)K$^2$. This led to the conclusion that $U_3Ni_3Sn_4$ has a FL ground state, with a crossover to NFL behaviour near $T \sim 0.5$ K. NFL-like temperature dependencies have also been observed in the magnetic and transport properties.

The electrical resistivity of single-crystalline $U_3Ni_3Sn_4$ under pressure ($p \leq 1.8$ GPa) was measured in the temperature range 0.3-300 K [10]. The residual resistivity, $r_0$, at ambient pressure equals $\sim 7$ μΩcm and $r_{RT}/r_0 = 55$, which shows that $U_3Ni_3Sn_4$ is a relatively clean material. The observed temperature variation of the resistivity at ambient pressure is typical for dense Kondo systems. The coherence temperature, $T_{coh}$, estimated by the temperature of the maximum in $dr/dT$, amounts to 20 K. The qualitative behavior of $r(T)$ does not change in the range of pressures applied. In the following we focus on the low-temperature behaviour. Under pressure the FL temperature interval increases. In Fig.4, we show $T_{FL}$ extracted by fitting the resistivity data to the expression $r = r_0 + AT^2$. $T_{FL}$ shows a strong variation with pressure. Within the theory of Rosch the linear dependence $T_{FL} \sim (p - p_c)$ is restricted to the immediate vicinity of the QCP, while at further distances to the QCP $T_{FL}$ shows a cross-over to $T_{FL} = a(p-p_c)^{1/2}$. The data obtained for our relatively clean sample of $U_3Ni_3Sn_4$ are consistent with the latter dependence. The solid line in Fig.4 represents the function $T_{FL} = a(p-p_c)^n$, with fit parameters $p_c = -0.04 \pm 0.04$ GPa, $n = 0.50 \pm 0.07$ and $a = 2.0 \pm 0.1$ KGPa$^{-n}$. Thus the analysis of the pressure variation of $T_{FL}$ within the magnetotransport theory of Rosch is consistent with $U_3Ni_3Sn_4$ being located close to an antiferromagnetic QCP, with the QCP located at a *negative* critical pressure of ~ -0.04 GPa. Fig.4 also shows that the coefficient $A$ of the $T^2$ term increases strongly upon approaching the QCP.

## 4  Conclusions

Resistivity measurements under pressure of the NFL correlated metals $U_2Pt_2In$ and $U_3Ni_3Sn_4$ show that Fermi liquid properties are rapidly recovered. A comparison of the data with the magnetotransport theory of Rosch provides evidence for an AF QCP at ambient pressure in $U_2Pt_2In$, and at a negative pressure of –0.04±0.04 GPa in $U_3Ni_3Sn_4$. Additional high-pressure experiments on samples with different amounts of disorder would be extremely useful to further investigate the scenario of a Hertz-Millis AF QCP as origin of the NFL properties in these materials.

**Acknowledgements**

The authors acknowlegde support within the EC-TMR and ESF-FERLIN programs.